\documentclass[a4paper,11pt]{article}
\pdfoutput=1 % if your are submitting a pdflatex (i.e. if you have
\usepackage{jinstpub} % for details on the use of the package, please
\usepackage{siunitx}
\usepackage{float}
\usepackage{subfig}
\usepackage{booktabs}
\usepackage{multirow}

\title{Characterisation of the Hamamatsu R12199-01 HA MOD photomultiplier tube for low temperature applications}% 
\author[a,1]{M.~A.~Unland~Elorrieta,%
\note{Corresponding author.}}
\author[a]{L.~Classen,}
\author[b]{J.~Reubelt,}
\author[c]{S.~Schmiemann,}
\author[b]{J.~Schneider}
\author[a]{and \mbox{A. Kappes}}

\affiliation[a]{Institut f\"ur Kernphysik, Westf\"alische Wilhelms-Universit\"at M\"unster}
\affiliation[b]{ECAP, Friedrich-Alexander-Universit\"at Erlangen-N\"urnberg}
\affiliation[c]{Institut f\"ur Theoretische Physik, Westf\"alische Wilhelms-Universit\"at M\"unster}

\emailAdd{m.unland@wwu.de}

\abstract{
We present a detailed characterisation of the new Hamamatsu R12199-01 HA MOD 3-inch photomultiplier tube (PMT) which is under consideration for the use in segmented optical modules of deep-ice neutrino detectors at the South Pole. Because of the significantly lower operation-temperature range compared to standard applications, a focus of our studies lies on the investigation of the temperature dependence of background characteristics (dark count rate, probability of correlated pulses), timing properties, gain and peak-to-valley ratio of this PMT type. In addition, the performance of the ``HA coating'' intended for background reduction was tested, as well as the influence of conductive objects near the photocathode like reflectors on the PMT noise rate. A low background rate is of particular importance as the deep ice at the South Pole features negligible optical background. We find that the new PMT type is well suited for the intended applications.
}


\begin{document}
\flushbottom
%\raggedbottom
\maketitle
\appendix

\section{\label{sec:introduction}Introduction}

Large-area photomultipliers are widely used in the field of neutrino detection. For instance, large-volume neutrino telescopes such as Antares \cite{Ageron_2011}, IceCube \cite{Aartsen_2017}, and the lake Baikal detector \cite{BDUNT}, but also smaller low-energy experiments such as Super Kamiokande \cite{Fukuda2003}, make use of $10+$ inch PMTs. However, with the KM3NeT detector \cite{Adri_n_Mart_nez_2016}, currently under construction in the Mediterranean Sea, and its optical sensors consisting of 31 three-inch tubes \cite{nim:a718:513}, fast and cost-effective small-area PMTs are being established as a viable alternative. The multi-PMT concept is now also foreseen for future extensions of the IceCube neutrino telescope \cite{Classen2017}.

The Hamamatsu type R12199-01 HA MOD PMT is based on the type R12199-02 PMT. The latter was derived by Hamamatsu for the KM3NeT experiment from the commercial flat-window type R6233 PMT with one of the main goals being improved timing properties. For IceCube, the R12199-02 was modified for the use in low-background applications with tight spacial constraints. To fit this usage profile, the PMT features a reduced tube length ($\SI{93}{mm}$ instead of \ $\SI{98}{mm}$) and a so-called ``HA-coating''. This is a conductive layer located on the outside surface of the tube surrounding the electron multiplier system and is electrically connected to the photocathode. This feature is designed to reduce and stabilise the dark noise rate of PMTs operated at negative high-voltage which is the foreseen polarity for the optical modules. 

The main characteristics of this new PMT model are expected to be comparable to those of the initial model. The latter was investigated thoroughly at room temperature by the KM3NeT collaboration using a dedicated PMT testing setup at the \emph{Erlangen Centre for Astroparticle Physics} (ECAP) in Erlangen \cite{km3netPaper, km3netsmall}. We tested 100 tubes of the new model using that very same setup. In addition, the temperature dependence of key parameters (gain, transit-time spread, pulse shape, peak-to-valley ratio, dark rate)\footnote{The temperature dependence of the quantum efficiency was not included, as it is expected to be small in the temperature range relevant for deep-ice detectors \cite{Singh1987}.} was investigated in detail for two specimen with a setup at the \emph{Institute for Nuclear Physics} in M\"unster in order to qualify them for deep-ice neutrino telescopes in Antarctica.

PMTs integrated in composite sensors are often equipped with reflectors or Winston Cones to enhance the light collection efficiency, e.g.\ in optical modules of IceCube extensions \cite{Classen2017}, KM3NeT optical modules \cite{nim:a718:513}, or the cameras of air Cherenkov telescopes \cite{Cornils2003,Lorenz2005}. The impact of such devices on the PMT dark noise as well as the effectiveness of the HA coating were also investigated.

\section{\label{sec:methods}Experimental setup and measurement method}

\begin{figure}[t]
    \centering
    \includegraphics[trim={0 8cm 1cm 0},clip, width=\textwidth]{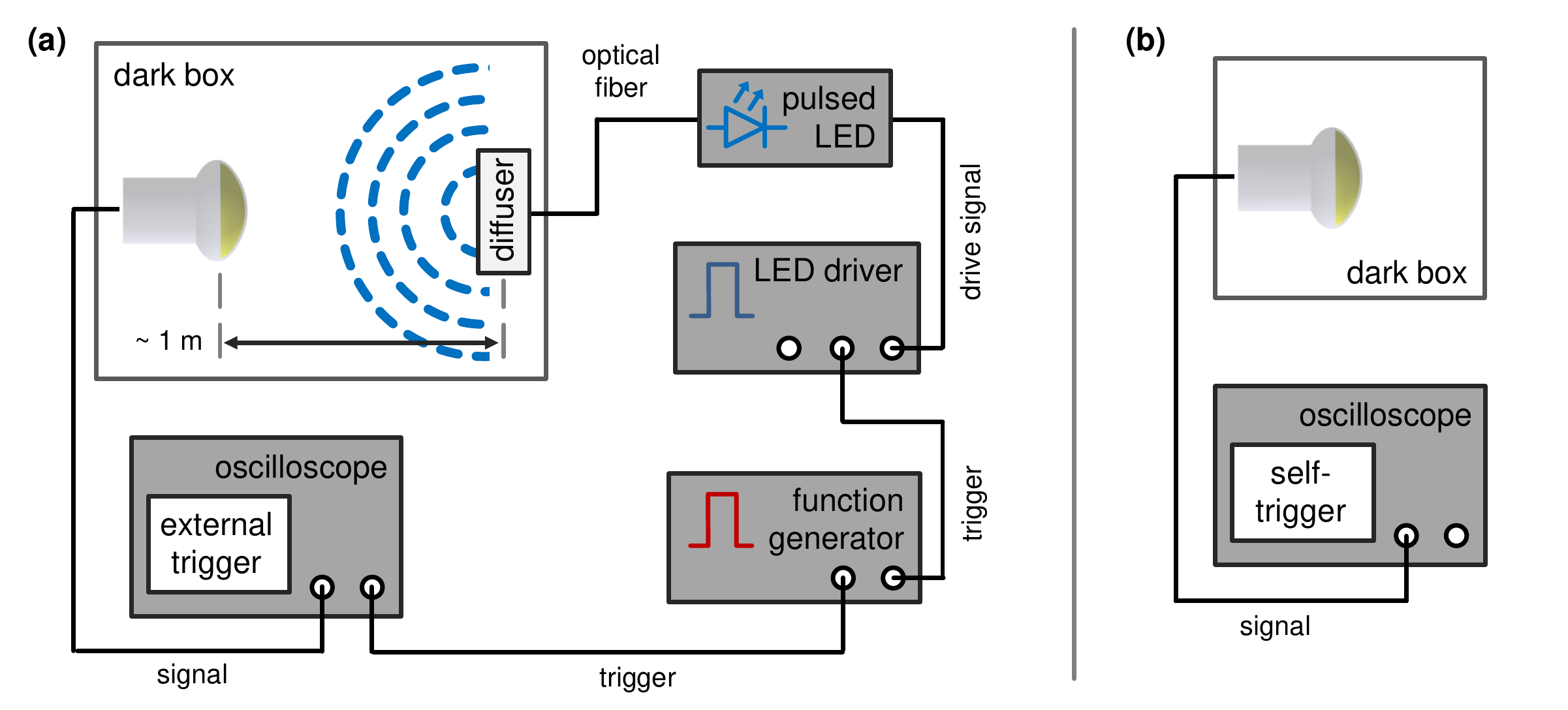}
    \caption{Schematic block diagram of the experimental setup for measurements of externally triggered waveforms in pulse mode (a) and of self-triggering measurements (b). Picture taken from \cite{LEW} and modified.}
    \label{secB:setup}
\end{figure}

All parameters discussed in this paper were measured with the PMTs operating in pulse mode. In this readout scheme, e.g.\ when employing an oscilloscope, the PMT charge output is fed to a resistor, which results in a voltage pulse. This way, individual pulses can be analysed in more detail, extracting information such as arrival time, amplitude and charge. Furthermore, either self-triggering on the PMT pulse or an external trigger (e.g.\ provided by a reference signal of a light source) can be used in this scheme.

A schematic diagram of both experimental setups is presented in Fig.~\ref{secB:setup}. The PMTs were placed in a light-tight box and were connected to an oscilloscope\footnote{Erlangen: LeCroy Waverunner 6100. M\"unster: PicoScope 6404C} and high-voltage supply\footnote{Erlangen: CAEN Mod. SY403 64 Channel High Voltage System. M\"unster: Iseg NHQ 226L}. For measurements of temperature dependencies, the dark box was placed inside a climatic chamber\footnote{CTS C-70/350}. Otherwise, the measurements were performed at room temperature ($\sim\SI{20}{\celsius}$) without any external control on the temperature. For all parameters except dark-noise rates, an external trigger was used. In this case, the PMT was illuminated with a pulsed LED\footnote{Erlangen: PicoQuant PDL 800B. M\"unster: wavelength of $\SI{385}{nm}$, mrongen -- Custom Picosecond light sources \cite{1748-0221-13-06-P06002}} that was triggered by a pulse generator\footnote{Erlangen: Agilent 33220A. M\"unster: mrongen -- Custom Picosecond light sources \cite{1748-0221-13-06-P06002}}, which also triggered the oscilloscope. In order to achieve a homogeneous illumination of the photocathode, a diffusor was placed in front of the light source at sufficient distance from the PMT.

\section{\label{sec:pulse}Pulse characteristics}

In this chapter we present the characteristics of the PMT read-out in pulse mode. In section~\ref{sec:charge_dist_gain}, charge spectrum characteristics are investigated as a function of temperature. Timing and pulse shape properties of the PMT are presented in section~\ref{sec:pulse_shape_timing}.

\subsection{Charge distribution and gain}
\label{sec:charge_dist_gain}

\begin{figure}[t]
\centering
\includegraphics[scale=0.995]{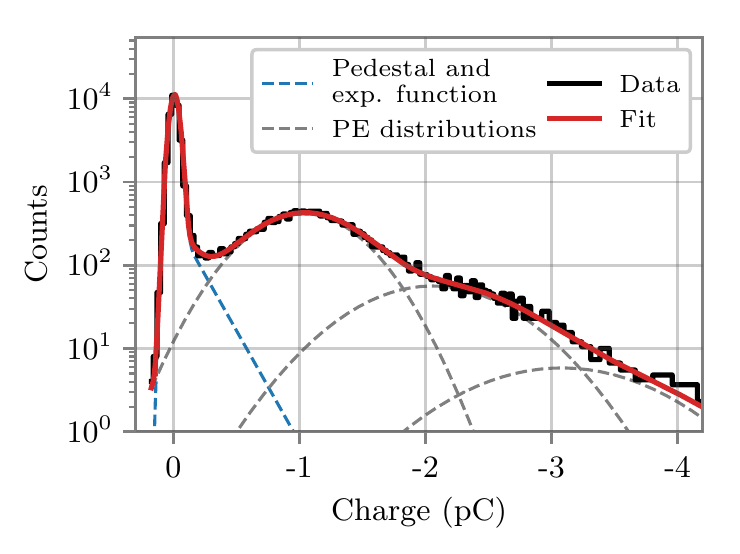}
\hfill
\includegraphics[scale=0.995]{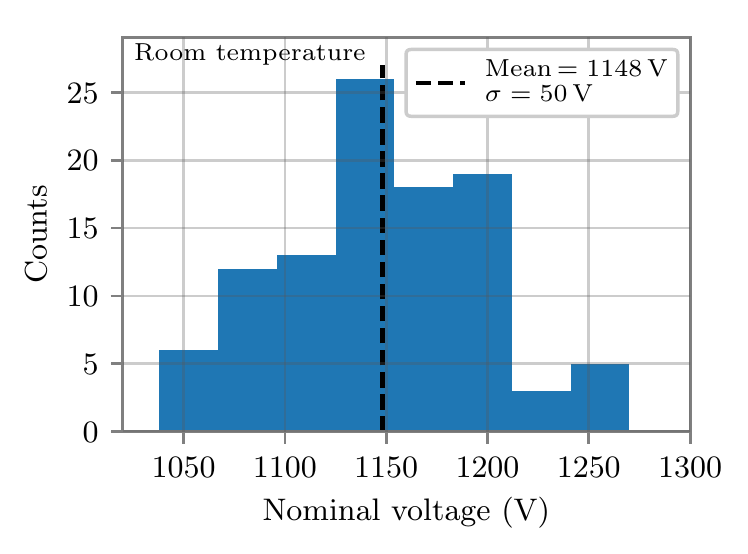}
\caption{\emph{Left}: Charge histogram of a PMT with its fit function in red. The blue dashed line marks the sum of the pedestal and the tail caused by under-amplified pulses. The grey dashed lines show the contribution from single to three PE. The data is fitted with the function introduced in eq. \ref{eq:GainFit}. \emph{Right}: Distribution of nominal voltages of the investigated PMTs for a gain of $5\times10^6$ at room temperature.}
\label{fig:SPE:charge-histo}
\end{figure}

The gain is the average number of electrons reaching the anode following the amplification of a single photoelectron (SPE). 
 To measure it, a light source illuminates the photocathode homogeneously and pulses are saved by triggering externally on the light source. The pulses are integrated over a time window and the resulting charge value is plotted into a histogram. Figure \ref{fig:SPE:charge-histo} shows a typical charge distribution, which features a pedestal originating from the baseline noise, followed by the peaks caused by photoelectrons (PE). The number of detected photons emitted by the light source is not constant but follows Poissonian statistics and thus the spectrum features contributions from SPEs and multiple PEs. The function used to model the charge spectrum in this work is based on \cite{Bellamy1994}, which presents a detailed description of the PMT calibration method utilised. The pedestal is described by a normal distribution with mean $Q_0$ and standard deviation $\sigma_0$. Also the charge distribution of $n$ PEs is modelled by a Gaussian distribution with a mean of $Q_n=Q_0+n\cdot Q_1$ and a standard deviation $\sigma_n = \sqrt{\sigma_0^2+n\cdot \sigma_1^2}$, where $Q_1$ is the average charge of an SPE and $\sigma_1^2$ its variance. There is a non-Gaussian component between the pedestal and the SPE peak, which we attribute to under-amplified pulses. These can be caused by photons creating a photoelectron at the first dynode or to inelastically back-scattered photoelectrons at the first dynode. This contribution is accounted for by an exponential function. Thus the distribution of the charge $q$ is described by
\begin{equation}
\label{eq:GainFit}
     f(q) = \textrm{e}^{-\mu}\cdot\Big[(1-P_u)\cdot G(q,Q_0,\sigma_0)+P_u \cdot\Theta(q-Q_0) \cdot \lambda \cdot\textrm{e}^{-\lambda \cdot (q-Q_0)}\Big]+\sum^\infty_{n=1}\frac{\mu^n \cdot \textrm{e}^{-\mu}}{n!}G(q,Q_n,\sigma_n),
\end{equation}
where $\Theta(x)$ is the step function, $G(x,\mu,\sigma) = 1/(\sqrt{2\pi}\sigma)\times\exp{(-(x-\mu)^2/2\sigma^2)}$ the normal distribution, $P_u$ the probability of under-amplified signals, $\lambda$ the coefficient of the exponential decay and $\mu$ the mean number of detected photoelectrons. The gain is calculated with $G={Q_1}/{e}$, where $e$ is the elementary charge. 
\begin{figure}[tb]
\includegraphics[scale=1]{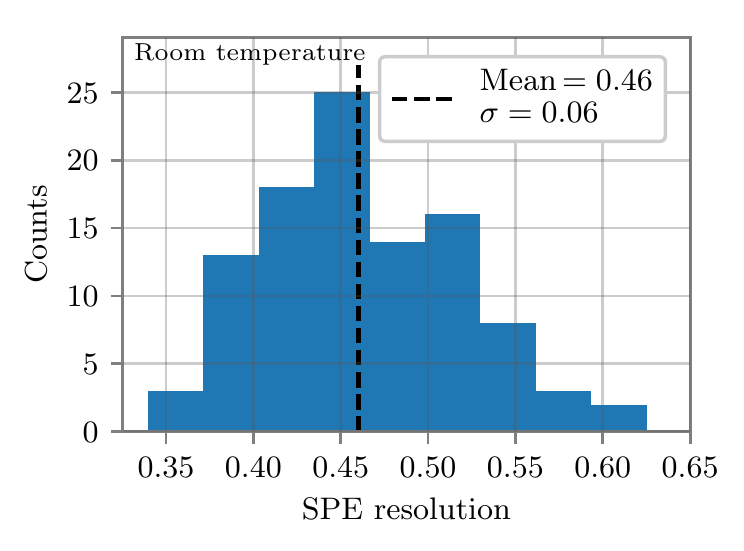}
\hfill
\includegraphics[scale=1]{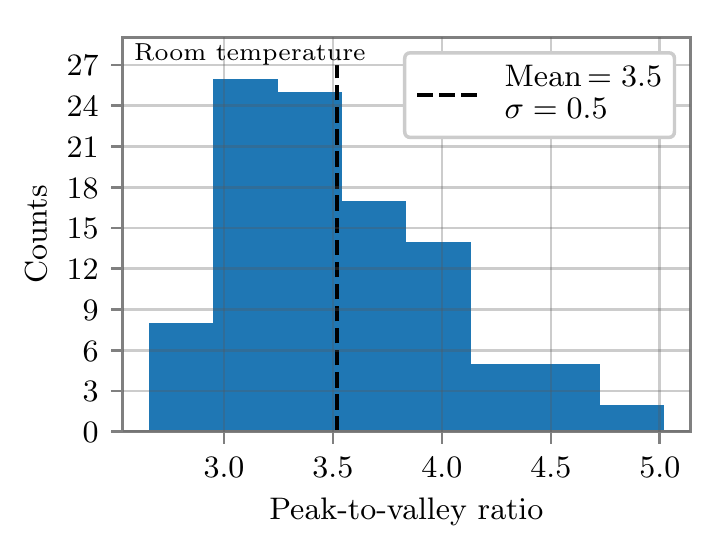}

\caption{Distribution of SPE resolution (left) and peak-to-valley ratio (right) for 102 PMTs at nominal voltage and at room temperature. The dashed black lines indicate the averages of the distributions.}
\label{fig:gen:p2vspr}
\end{figure}

The gain of every PMT is measured at different voltages in order to determine its nominal voltage defined here as the voltage $V$ at which the PMT features a gain of $G=5\times10^6$. The selection of this nominal gain is a compromise between improving the SPE and time resolution of the PMT and keeping its dark rate low \cite{hamamatsuBook}. For the gain calibration, the data is fitted with a linear function in double logarithmic scale $\log{(G)}=a\cdot \log{(V)}+b$. The nominal voltage distribution of the $102$ PMTs is shown in Fig.~\ref{fig:SPE:charge-histo}. The average of the distribution is $\SI{1148}{V}$ with a standard deviation of $\SI{50}{V}$.

\begin{figure}[t]
\includegraphics[scale=1]{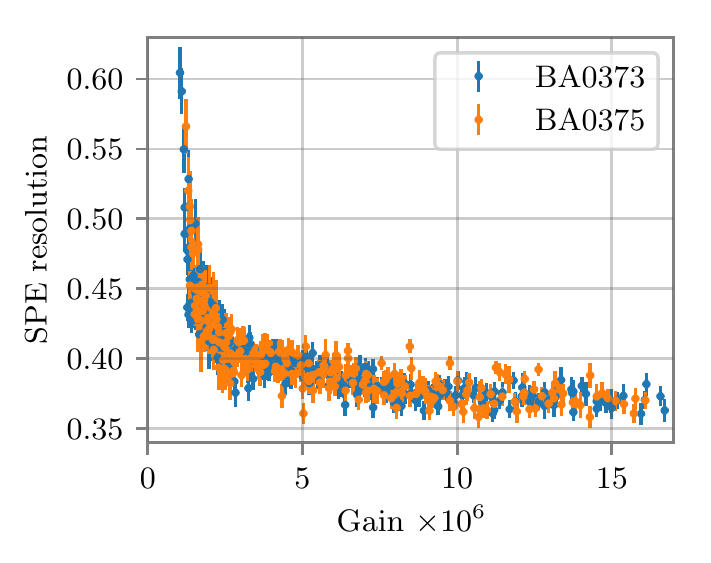}
\hfill
\includegraphics[scale=1]{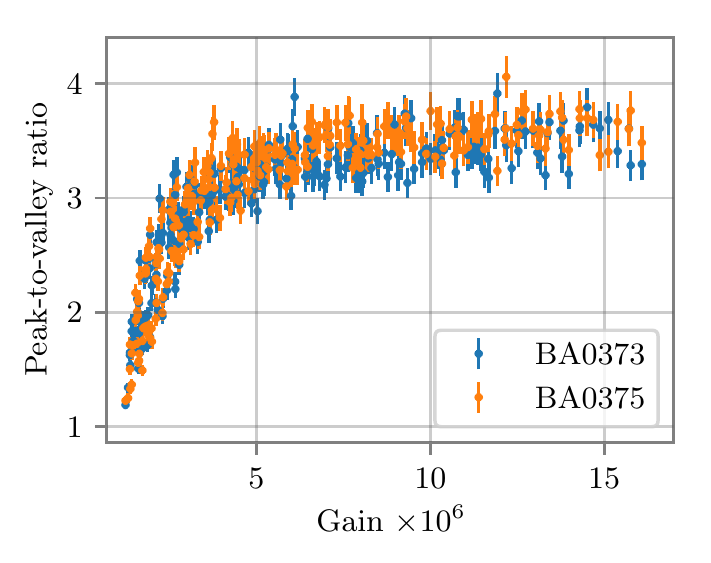}

\caption{SPE resolution (left) and peak-to-valley ratio (right) as a function of gain.}
\label{fig:gen:p2v}
\end{figure}

Several gain-related parameters were obtained from the charge spectrum, such as the SPE resolution ${\sigma_1}/{Q_1}$ and the peak-to-valley ratio. The latter is defined as the ratio between the minimum of the valley between the pedestal and the SPE peak, and the maximum of the SPE peak. These parameters are related to the capability of the PMT to distinguish SPE and noise signals. The distributions of the SPE resolution and the peak-to-valley ratio are shown in Fig.~\ref{fig:gen:p2vspr} for the 102 PMTs at nominal voltage. The average SPE resolution is $\SI{0.46}{}$ with a standard deviation of $0.06$, while the mean for the peak-to-valley ratio is $\SI{3.5}{}$ with a standard deviation of $0.5$. Both values improve with increasing gain, as can be seen in Fig.~\ref{fig:gen:p2v}.

\begin{figure}[tb]
\includegraphics[scale=1]{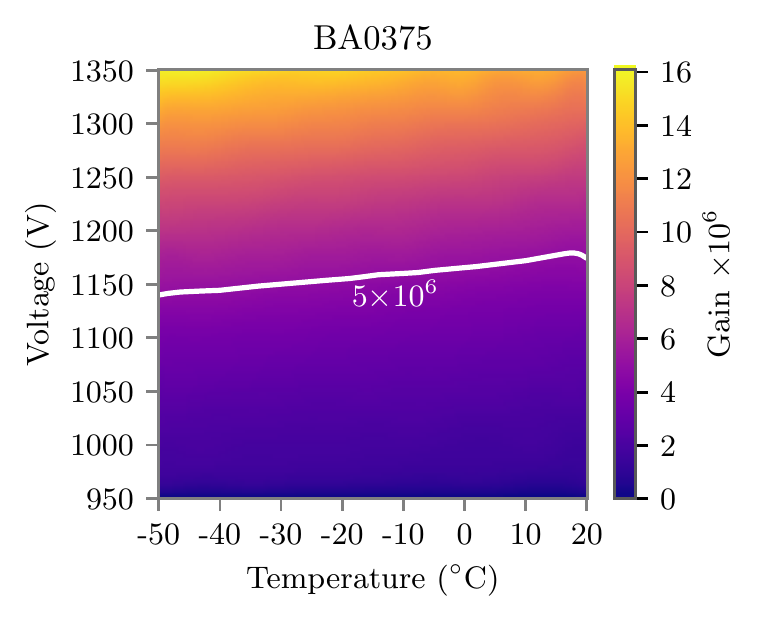}
\hfill
\includegraphics[scale=1]{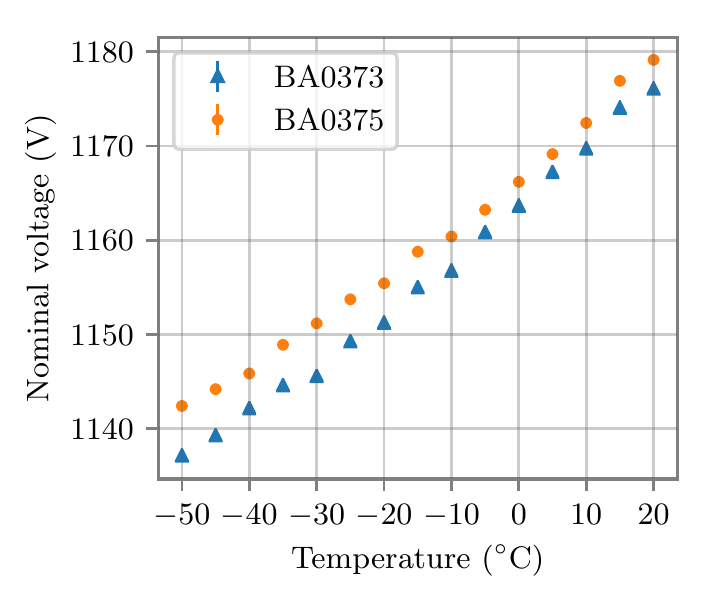}
\caption{\emph{Left:} Gain as a function of temperature and voltage (interpolated). The gain was determined for voltages between $\SI{950}{V}$ and $\SI{1350}{V}$ in steps of $\SI{25}{V}$ and at temperatures between $\SI{-50}{\celsius}$ and $\SI{+20}{\celsius}$ in steps of $\SI{5}{\celsius}$. \emph{Right:} Increase of nominal voltage due to the decrease in gain with temperature. Error bars are smaller than data markers.}
\label{fig:gen:gain_temp}
\end{figure}

\par
\begin{figure}[t]
    \centering
    \includegraphics[scale=1.0]{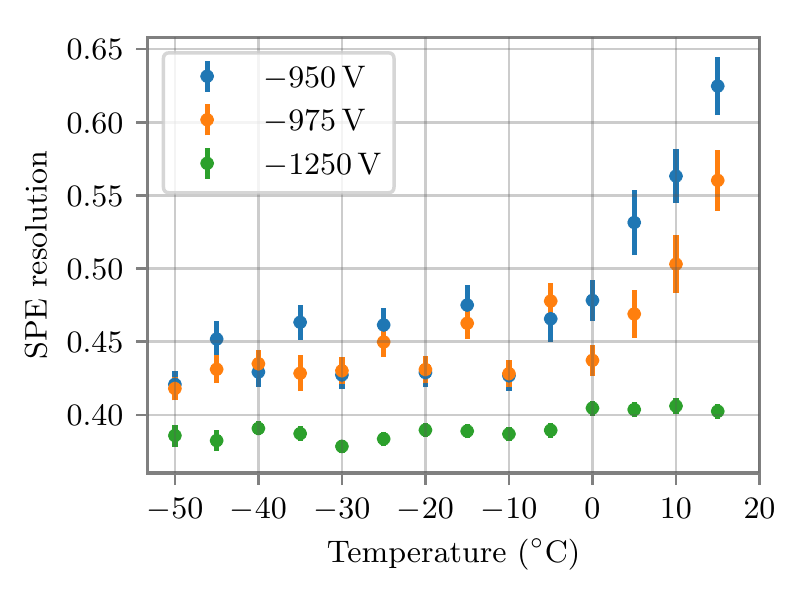}
    \caption{SPE resolution of PMT BA0375 as a function of temperature at three different voltages.}
    \label{fig:gen:SPEResoV}
\end{figure}

 In order to investigate the temperature dependence of the gain, two tubes were placed inside a climatic chamber and the gain calibration was carried out at temperatures between $\SI{-50}{\celsius}$ and $\SI{20}{\celsius}$ (hereinafter the PMTs are referred to by their serial numbers BA0373 and BA0375). The gain of BA$0375$ is shown in Fig.~\ref{fig:gen:gain_temp} as a function of the voltage and the temperature. In agreement with other studies \cite{wrightgaintemp, Barrow:2016doe}, the average SPE charge decreases with the temperature. The mean temperature coefficient of BA$0373$ and BA$0375$ are $\SI{-0.368\pm0.004}{\% \celsius^{-1}}$ and $\SI{-0.398\pm0.005}{\% \celsius^{-1}}$, respectively. Hence, cooling by $\Delta T = \SI{70}{\celsius}$ results in an increased gain of $26\%$ and $28\%$, respectively. Similar numbers have been reported in the cited publications, nevertheless, a clear explanation for this effect is not given. The measured charge can increase either as the result of a decreasing resistivity of the signal cable, the base or the PMT, or of an actual increase of the PMT gain due to a higher secondary emission coefficient of the dynodes.
If the decrease in SPE charge with temperature would only be due to an increase in the resistance, the SPE resolution should not vary with temperature. Measurements of the SPE resolution at different temperatures are shown in Fig.~\ref{fig:gen:SPEResoV} for voltages of $\SI{-950}{V}$,$\SI{-975}{V}$ and $\SI{-1250}{V}$. At $\SI{-1250}{V}$ the PMT has a gain where the SPE resolution remains fairly constant with small gain variations (see Fig.~\ref{fig:gen:p2v}). Therefore, the SPE resolution does not vary strongly with temperature at this voltage. For lower voltages, the resolution decreases from $\sim 0.6$ at room temperature to $\sim 0.45$ at $\SI{-50}{\celsius}$. This favours the hypothesis of an increase of the secondary emission coefficient, although a contribution from a changed resistance of the setup cannot be excluded. Nevertheless, this effect reduces the nominal voltage when the PMT is cooled as shown in the right plot of Fig.~\ref{fig:gen:gain_temp}. This demonstrates the necessity for an in-situ calibration of the PMTs.

\subsection{Pulse shape and timing properties}
\label{sec:pulse_shape_timing}

The time resolution of the PMT is determined by the pulse shape of an SPE signal and the transit time spread (TTS). The latter is the variation of the signal transit time (TT), i.e.\ the time between the photon absorption in the photocathode and the charge collection at the anode. The pulse shape determines the shortest time interval between two pulses that a single PMT can resolve and the TTS is the detection time uncertainty of an SPE.

\begin{figure}[t]
\includegraphics[scale=1, trim=0 -0.8cm 0 0]{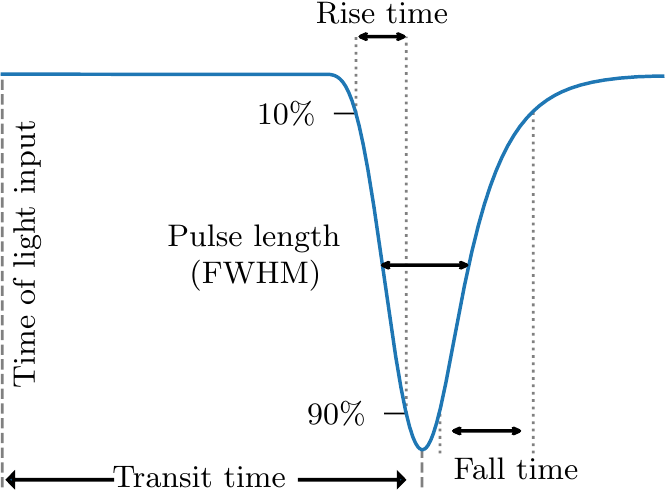}
\hfill
\includegraphics[scale=1]{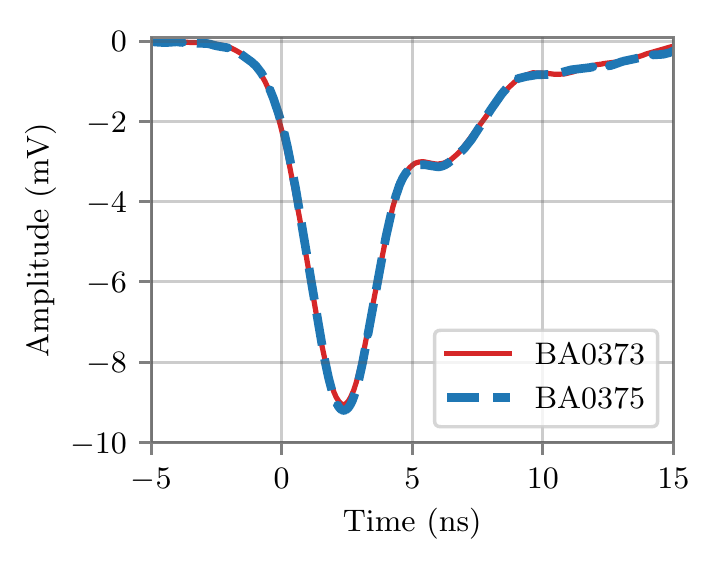}

\caption{\emph{Left}: Visualisation of shape and timing parameters of PMT pulses. Picture taken from \cite{martin}. \emph{Right}: Average of $5000$ pulses of PMT BA$0373$ (red) and BA$0375$ (blue, dashed) at a temperature of $\SI{20}{\celsius}$.}
\label{fig:SPE:time_av_pulse}
\end{figure}

The right-hand side of Fig.~\ref{fig:SPE:time_av_pulse} shows the average of $5000$ pulses of two PMTs, both operated at a gain of $5\times10^6$. The left-hand side of Fig.~\ref{fig:SPE:time_av_pulse} shows the shape parameters: the pulse length, given by its FWHM, the rise time (RT) and the fall time (FT). The rise (fall) time is defined as the time needed for the pulse to rise (fall) from $\SI{10}{\percent}$ ($\SI{90}{\percent}$) to $\SI{90}{\percent}$ ($\SI{10}{\percent}$) of the pulse amplitude. These parameters were measured for $\sim4500$ pulses at different temperatures between $\SI{-50}{\celsius}$ and $\SI{20}{\celsius}$. Also the relative transit time is calculated storing the time of the pulse minimum. Since the TTS is determined with SPEs, the light source was set at such intensity that the average number of detected photons was $<\SI{0.1}{PE}$ and hence, the probability of multiple PE pulses $<\SI{0.5}{\percent}$. The voltage was adjusted as a function of temperature in order to keep the PMTs at nominal gain.

\begin{figure}[t]
\centering
\includegraphics[scale=1]{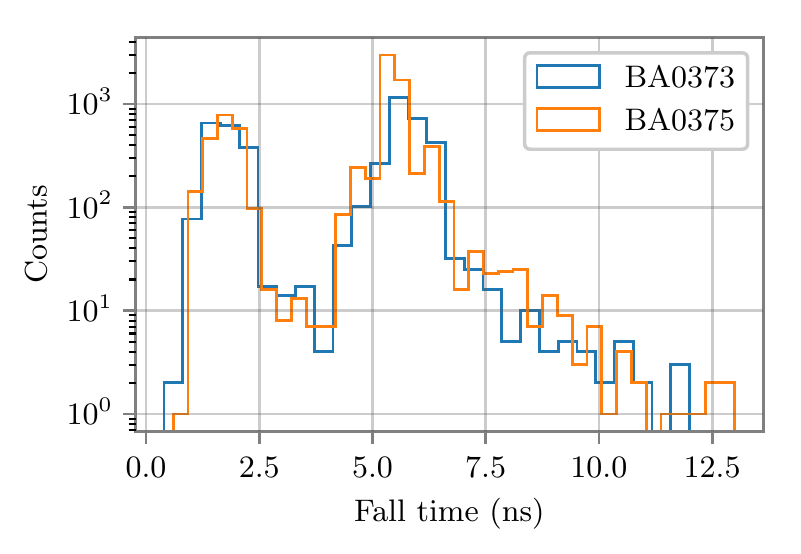}
\caption{Fall time of about $4500$ SPE pulses at a temperature of $\SI{20}{\celsius}$.}
     \label{fig:SPE:FThisto}
\end{figure}

\begin{figure}[tb]
\centering
\includegraphics[scale=1]{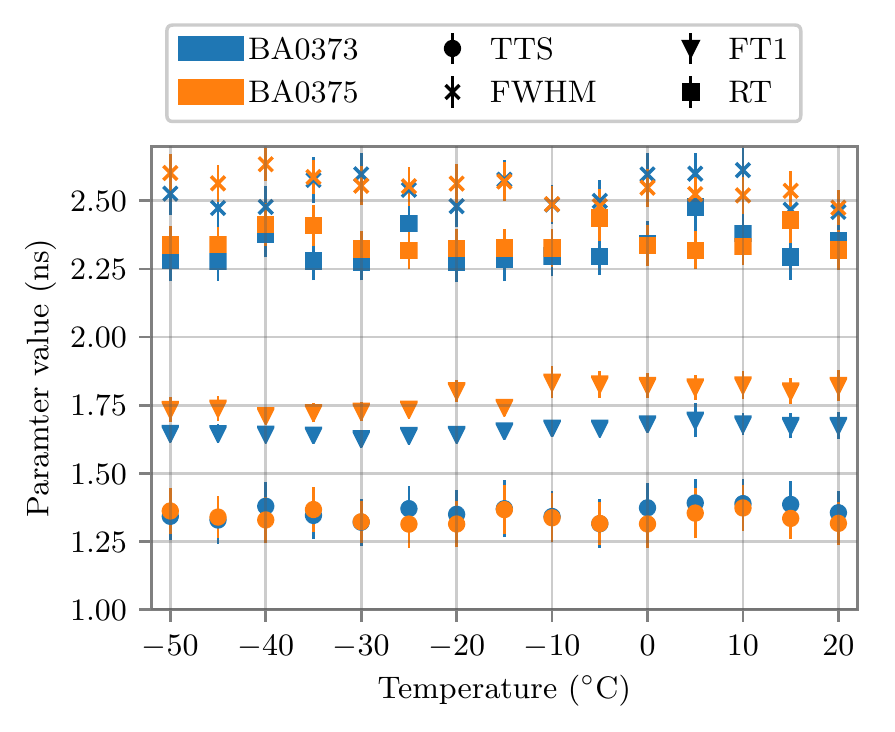}
\caption{Transit-time spread (TTS), fall time (FT), rise time (RT) and pulse width (FWHM) of PMTs BA$0373$ and BA$0375$ as a function of temperature at nominal voltage.}
\label{fig:SPE:timing_temperature}
\end{figure}

 The voltage divider used in the measurements causes a ringing as visible in Fig.~\ref{fig:SPE:time_av_pulse} right. Figure~\ref{fig:SPE:FThisto} shows the distribution of the measured fall times at $\SI{20}{\celsius}$. Since the pulse shape varies, some pulses feature a ringing larger than $\SI{10}{\percent}$ of the pulse amplitude and others do not. Therefore, two peaks can be identified at $\sim\SI{1.7}{ns}$ and $\sim\SI{5.5}{ns}$.
In the following, FT1 refers to the average fall time from pulses with low amplitude (or absent) ringing and FT2 from those with ringing amplitude larger than $\SI{10}{\percent}$ of the pulse amplitude. The averages of the timing parameters are shown in Fig.~\ref{fig:SPE:timing_temperature} as a function of temperature at nominal voltage. They remain practically constant throughout the investigated temperature range. The average value of each parameter can be found in Table \ref{tab:SPE:timing_temperature}.

\begin{table}[t]
\caption{Mean values of timing parameters for PMTs BA$0373$ and BA$0375$.}
\label{tab:SPE:timing_temperature}
\centering

\begin{tabular}{@{}lll@{}}
\toprule
          & BA$0373$        & BA$0375$        \\ \midrule
RT (ns)   & $\SI{2.319\pm0.016}{}$ & $\SI{2.348\pm0.011}{}$ \\
FT1 (ns)  & $\SI{1.653\pm0.005}{}$ & $\SI{1.757\pm0.012}{}$ \\
FT2 (ns)  & $\SI{5.561\pm0.015}{}$ & $\SI{5.714\pm0.011}{}$ \\
FWHM (ns) & $\SI{2.520\pm0.015}{}$ & $\SI{2.547\pm0.013}{}$ \\
TTS (ns)  & $\SI{1.357\pm0.007}{}$ & $\SI{1.336\pm0.006}{}$ \\ \bottomrule
\end{tabular}
\end{table}

\vspace{\baselineskip}

\begin{figure}[tb]
\includegraphics[scale=1]{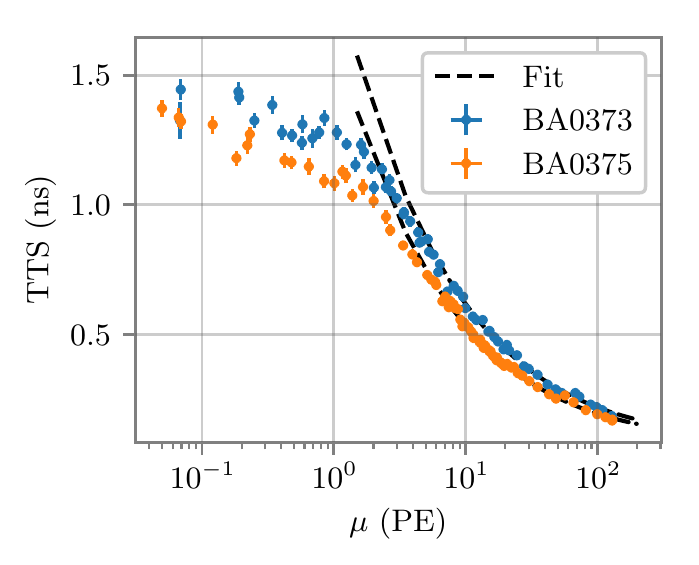}
\hfill
\includegraphics[scale=1]{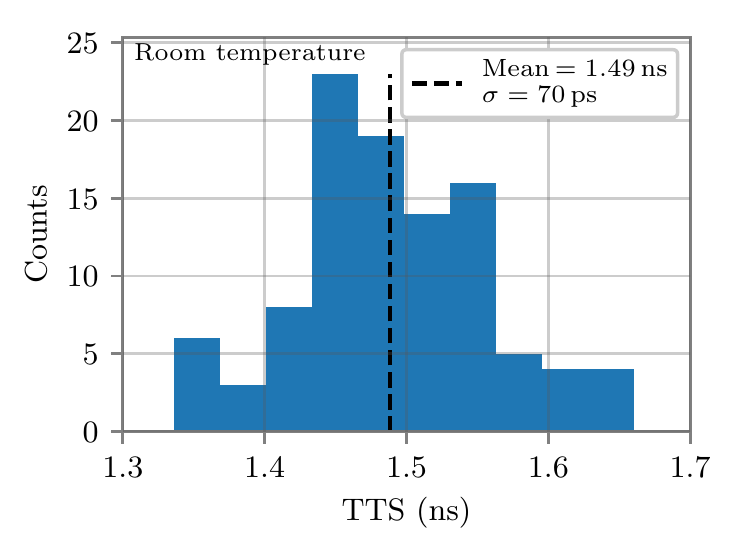}
\caption{\emph{Left}: Transit-time spread as a function of average PMT pulse charge $\mu$ in PE. For $\mu>5$, the curves were fitted with the function $f(\mu)=(a/\sqrt{\mu}+b)$, shown with dashed black lines. \emph{Right}: TTS distribution of $102$ PMTs for SPEs. The black dashed line indicates the mean of the distribution.}
\label{fig:SPE:tts_vs_mu}
\end{figure}

The TTSs of the PMTs BA$0373$ and BA$0375$ were also measured as a function of the light intensity (Fig.~\ref{fig:SPE:tts_vs_mu} left). Since the number of detected photons follows Poisson statistics, the TTS does not change much at low light levels (mean number of photons $\mu < \SI{0.3}{PE}$), as almost exclusively SPEs are being measured. With increasing light intensity the TTS decreases and above $\mu\sim \SI{5}{PE}$ it is proportional to ${1}/{\sqrt{\mu}}$ (black dashed line), consistent with other publications \cite{Huang:2014yfa, Liao:2017cdh}. It should be noted that this is not an intrinsic PMT property, but an expected result from statistics. Assuming that the TT of different photoelectrons are independent and uncorrelated featuring a variance $\rm{Var}(\rm{TT})= \sigma^{2}$, the variance of the mean transit time $\overline{\textrm{TT}}$ of $n$ photoelectrons is
\begin{equation}
\textrm{Var}\big(\overline{\textrm{TT}}\big)=\textrm{Var}\Big(\frac{1}{n}\sum^{n}_{i=1}\textrm{TT}_i\Big)=\frac{1}{n^2}\sum^{n}_{i=1}\textrm{Var}(\textrm{TT}_i)= \frac{\sigma^2}{n}.  
\end{equation}
The TTS distribution for SPEs of $102$ PMTs is shown on the right side of Fig.~\ref{fig:SPE:tts_vs_mu}. All PMTs feature similar values with a standard deviation of only $\SI{70}{ps}$ and a mean TTS of $\SI{1.49}{ns}$.

\section{\label{sec:Background}Background}
A photomultiplier always produces a measurable signal even in total darkness. If operated in pulse mode, this output is referred to as dark rate. The background can be classified into random and correlated noise. All pulses have a certain probability to be replaced by early or delayed pulses which feature a different timing than expected from transit time and TTS. Furthermore, pulses can be succeeded by afterpulses. The first part of this section will present the temperature dependence of the dark rate and the influence of external electromagnetic fields. In the second part the probabilities for correlated pulses are measured and their temperature dependence is investigated.

\subsection{Dark rate}
\begin{figure}[t]
    \centering
    \includegraphics[scale=1.0]{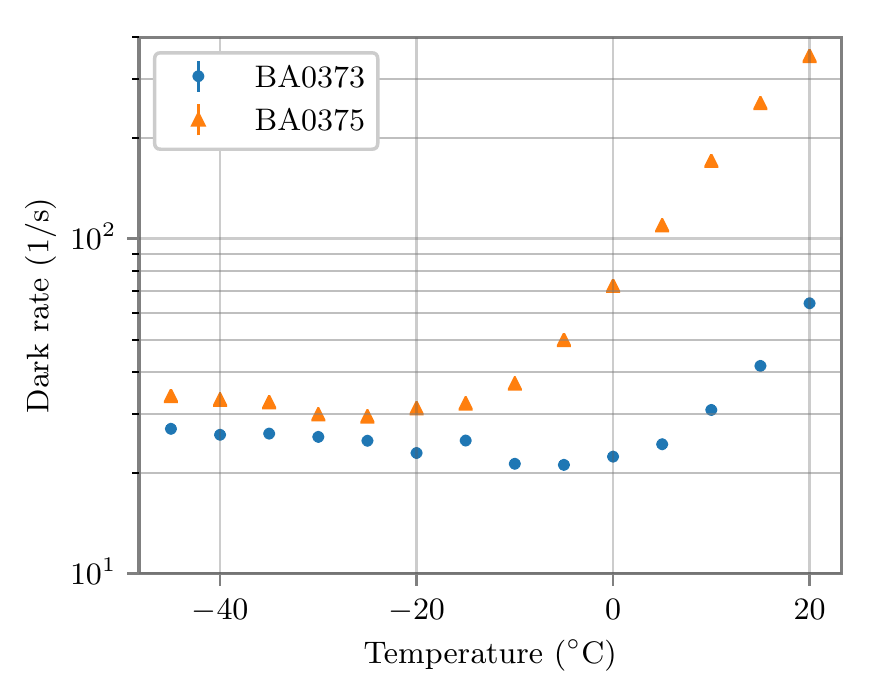}
    \caption{Dark rate as a function of temperature for PMTs BA$0373$ and BA$0375$. The voltage was adjusted in order to keep the PMTs at nominal gain. Error bars are smaller than data markers.}
    \label{DR:fig:drtemp}
\end{figure}

Light exposure during storage of the PMT can temporarily increase the dark rate. It can take between a couple of hours to days until the PMT recovers to a normal dark rate level. Each PMT was kept in dark at least $\SI{12}{\hour}$ before the measurements were started. The time $t$ needed for detecting $n$ events surpassing the trigger level of $\sim0.3$PE was measured. The dark rate $R$ was determined as $R=n\cdot t^{-1}$. This counting method has a dead time on the order of $\SI{1}{\mu s}$ between pulses. This is caused by the dead time of the data acquisition system\footnote{Picoscope 6404C} due to the re-arming of the trigger.

Figure~\ref{DR:fig:drtemp} depicts the dark rate of PMTs BA$0373$ and BA$0375$ as a function of temperature. The voltage was adjusted as a function of temperature in order to keep the PMTs at nominal gain. At room temperature, the background is dominated by electrons released in the photocathode due to thermionic emission \cite{hamamatsuBook}. Its magnitude is quite different for both PMTs with $\SI{64}{s^{-1}}$ for BA0373 and $\SI{351}{s^{-1}}$ for BA0375, respectively. This difference is significantly reduced as the contribution of the thermionic emission decreases towards lower temperatures. At these temperatures, field emission and scintillation photons produced by radioactive decays and interaction of cosmic particles inside the envelope glass are the most important background sources \cite{hamamatsuBook}. The average rate between $\SI{-45}{\celsius}$ and $\SI{-15}{\celsius}$ is  $\SI{25.6\pm0.6}{s^{-1}}$ and $\SI{31.7\pm0.8}{s^{-1}}$ for the PMTs BA0373 and BA0375, respectively. In this temperature region there is a slight increase of the rate with decreasing temperature, which is probably caused by the rise of the scintillation yield of the glass \cite{martin}.

Correlated noise (i.e.\ afterpulsing) is not completely included in this measurement due to the aforementioned dead time of the data acquisition system. Therefore, the real background rate of the PMTs is slightly higher than the measured one.

\subsubsection*{Influence of external components on dark rate}

\begin{figure}[t]
\includegraphics[width=\textwidth]{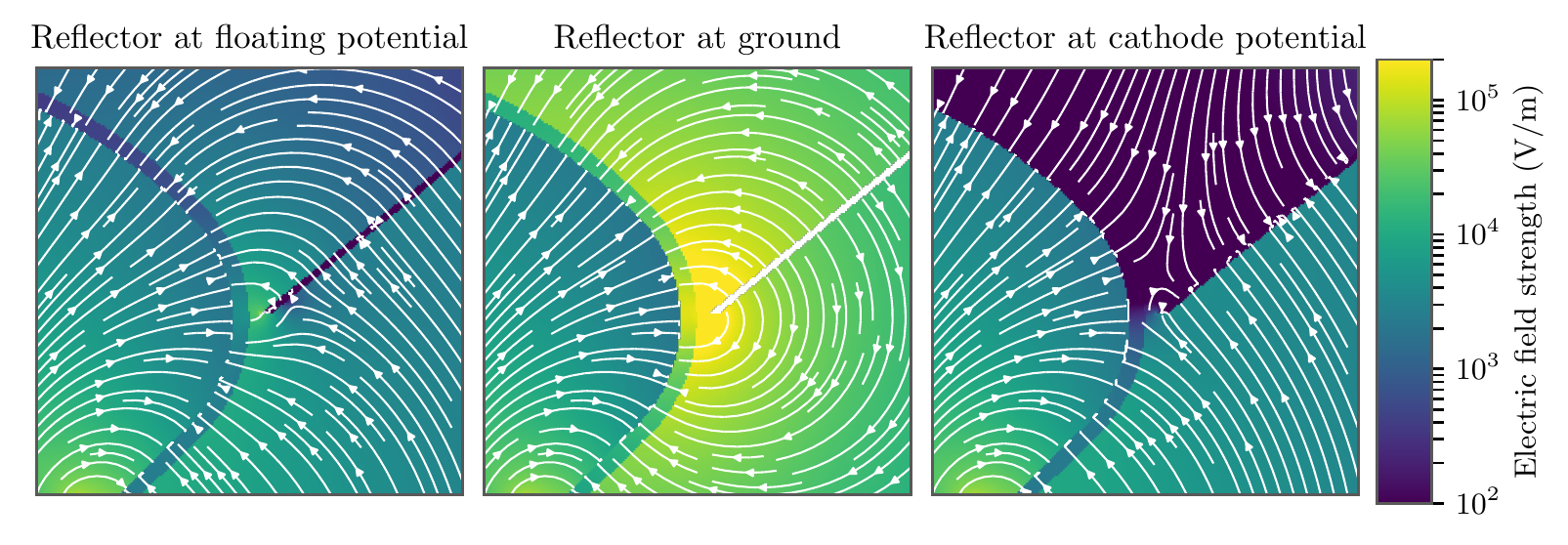}\\[-3mm]
\includegraphics[width=\textwidth]{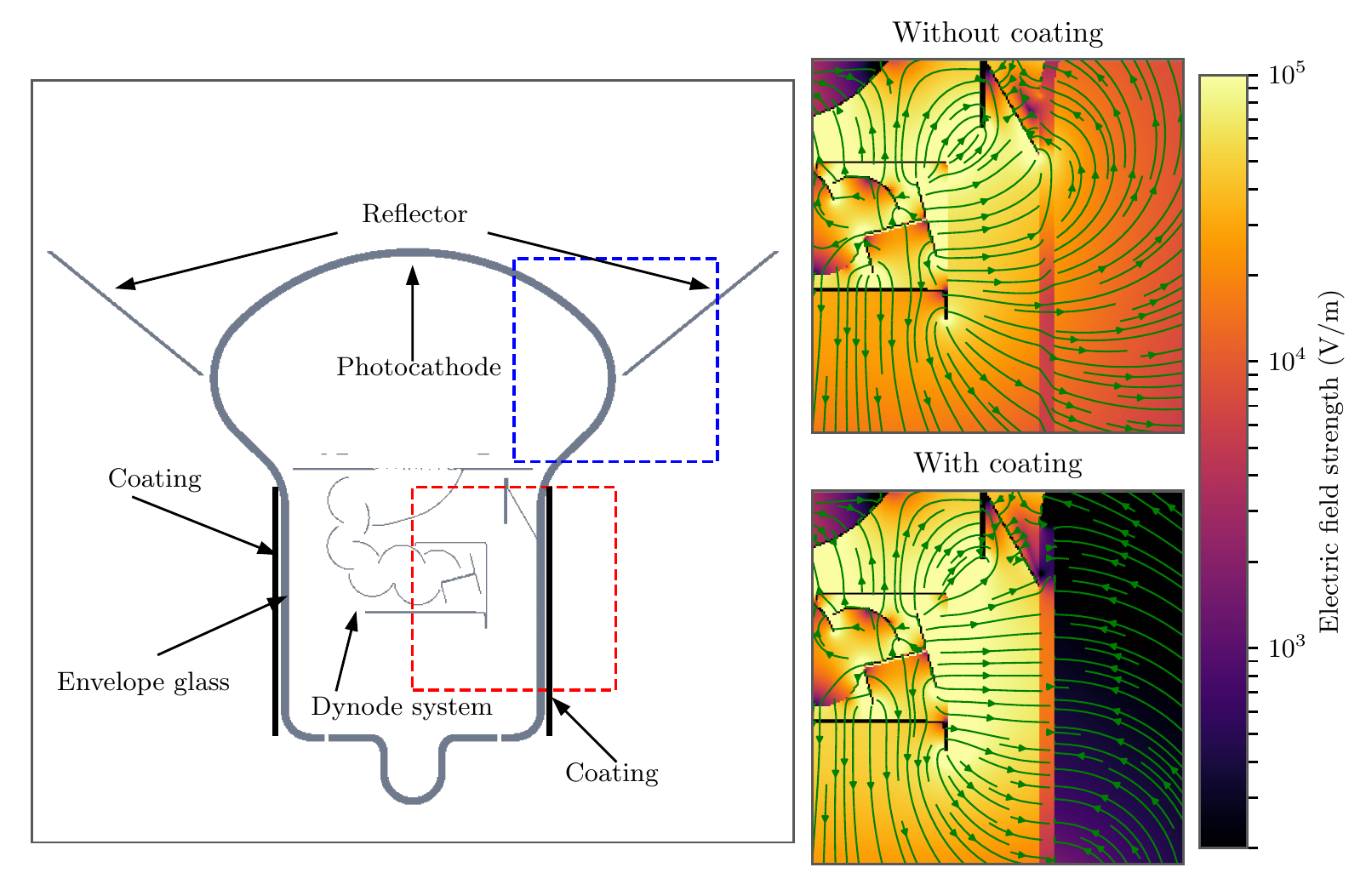}

\caption{\emph{Lower left}: Cross-section of the simulated geometry. The blue dashed box indicates the region of the plots in the top row and the red box those of the two plots on the lower right. Coating thickness was increased for the sake of clarity. \emph{Top row:} Finite-element simulation of the electric field around the PMT entrance window with the reflector on floating potential (left), ground (middle) and photocathode potential (right). \emph{Lower right}: Electric field around the tube's envelope with and without a coating at photocathode potential. The colour scale indicates the electric field strength in $\SI{}{\volt/\meter}$ and the arrows indicate the direction of the field in the plane.}
\label{fig:dr:comsol}
\end{figure}

PMTs operated at negative high-voltage feature a higher dark rate due to the influence of external electrical fields \cite{Wright2017}. Thus, it is necessary to avoid the placement of conductive objects like reflectors close to the tube. In particular grounded conductive objects should not be placed near the PMT. A usual method for mitigating this problem is the usage of a conductive coating around the PMT connected to the photocathode potential. Another approach investigated in \cite{pmtcover} is to coat the PMT with insulating varnish, which prevents possible discharges between surrounding objects and the tube. Nevertheless, a coating does not protect the photocathode, which is in particular sensitive to electromagnetic fields \cite{Wright2017}.

The influence of electric potentials close to the Hamamatsu PMT has been investigated with a finite element simulation using COMSOL Multiphysics$^{\circledR}$\footnote{COMSOL Multiphysics v. 5.2. www.comsol.com. COMSOL AB, Stockholm, Sweden.}. All components of the PMT were modeled with Autodesk Inventor$^{\circledR}$\footnote{Autodesk Inventor 2018. www.autodesk.eu/products/inventor/. Autodesk, Inc., California, U.S.} using realistic dimensions. The bottom left part of Fig.~\ref{fig:dr:comsol} shows a cross-section of the simulated geometry. Electric potentials have been assigned to the different parts of the PMT corresponding to a cathode-anode voltage of $\SI{-1300}{V}$. The simulation results shown in Fig.~\ref{fig:dr:comsol} are explained and discussed in the following sections.

\paragraph{HA coating:}
Any conductive object near ground potential in the vicinity of a PMT operated at negative high-voltage will alter the path of photoelectrons and charged secondaries. The electrons may hit the glass envelope and produce background PMT pulses either by releasing electrons from alkali metals settled on the walls during photo-activation of the PMT or via luminescence \cite{Wright2017}. As previously described, the Hamamatsu PMT features a coating around the tube consisting of an inner conductive layer at photocathode potential surrounded by an insulating layer (see PMT sketch in Fig.~\ref{fig:dr:comsol}). The two pictures at the lower right of Fig.~\ref{fig:dr:comsol} show the resulting electric field configuration near the dynode system and envelope glass for a PMT with and without such coating. The utilisation of the coating increases the electric field strength between the dynode system and the glass by $\sim\SI{80}{\percent}$. This results in an increased repulsion of the electrons from the glass towards the anode and therefore a reduction of the background rate. In addition the coating protects the volume of the dynode structure from external electric fields.

\begin{figure}[tb]
\includegraphics[scale=1]{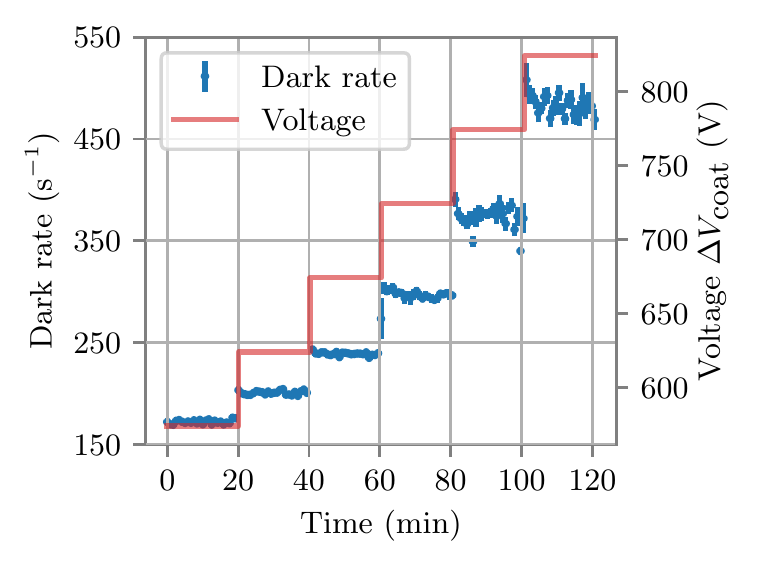}
\hfill
\includegraphics[scale=1]{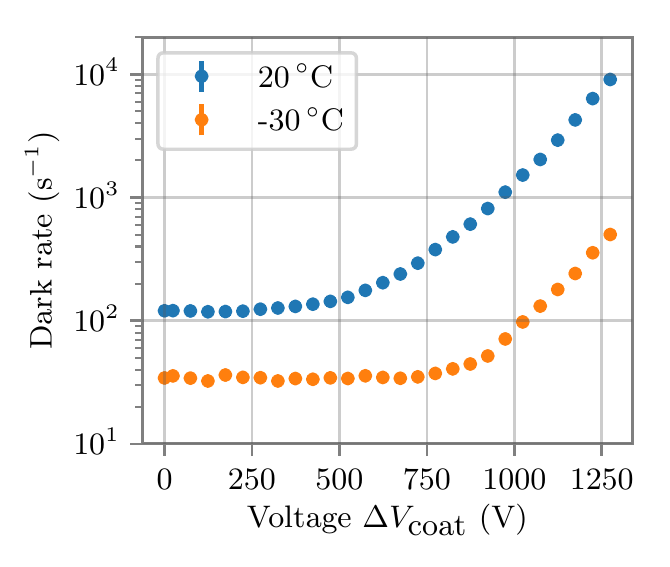}
\caption{Dark rate of a PMT with its tube coated by an aluminium band at different voltages with respect to the photocathode. Error bars are smaller than the markers. \emph{Left:} Voltage and dark rate as a function of time at room temperature. \emph{Right:} Dark rate as function of voltage.}
\label{fig:dr:coatingratechangetime}
\end{figure}

In order to further investigate the effects of this coating empirically, an aluminium adhesive tape\footnote{Duck$^{\circledR}$ Aluminium Tape} was attached to the tube of an uncoated Hamamatsu PMT. This tape covered the same region that the HA coating would envelope. It was connected to a different channel of the PMT power-supply in order to obtain a common ground. The left side of Fig.~\ref{fig:dr:coatingratechangetime} presents the change of the dark rate increasing the tape voltage in intervals of 20 minutes. The dark rate increases step-wise with the potential difference between the tape and the photocathode $\Delta V_{\textrm{coat}}$. This clearly demonstrates the adverse effects of external fields on an unprotected dynode structure. The right side of Fig.~\ref{fig:dr:coatingratechangetime} shows the dark rate as a function of $\Delta V_{\textrm{coat}}$ for the PMT at room temperature and $\SI{-30}{\celsius}$. In both cases the rate does not seem to be affected at voltages near the photocathode potential, but it quickly rises for larger $\Delta V_{\textrm{coat}}$, reaching a maximum of $9000\,s^{-1}$ ($\SI{20}{\celsius}$) and $\SI{510}{s^{-1}}$ ($\SI{-30}{\celsius}$) with the aluminium band at ground.
This is in agreement with results from \cite{Wright2017}, although there the measurement was performed only at room temperature. This observed behaviour is compatible with the luminescence hypothesis: at larger potential differences, more electrons are deflected towards the glass envelope, increasing the total rate. Nevertheless, the exponential increase of the rate starts at different potential differences for room temperature and $\SI{-30}{\celsius}$. This temperature dependent difference may suggest a more complex effect. Further studies regarding time correlation of this background may be pertinent in order to better understand this behaviour. It is clear, however, that if a low background is required, the high-voltage coating should be an effective method of protecting the PMT against external fields. 

\paragraph{Reflector:}
Aluminium reflectors are commonly used to enhance the effective area of PMTs (e.g.\ in KM3NeT \cite{nim:a718:513}, MAGIC \cite{Lorenz2005}, H.E.S.S.\ \cite{Cornils2003}). However, being conductive objects their utilisation may result in undesirable effects since they are placed at the PMT window near the photocathode, where the tube can not be protected by a high-voltage coating. The top row of Fig.~\ref{fig:dr:comsol} shows the electric field strength near the photocathode and the reflector for three scenarios. The left plot shows the reflector at floating potential. This leads to high field strengths in the PMT envelope glass, with a maximum of $\SI{4.6}{kV/m}$ in the region of smallest distance of the reflector to the PMT ($\SI{1}{mm}$ distance). If the reflector is at ground potential (middle plot) the field strength increases dramatically to $\SI{0.14}{MV/m}$ in this region. Areas further away from the reflector also feature relatively large values if the reflector is either at ground or at floating potential. At the glass surface near the centre of the photocathode, the field strength amounts to $\SI{12}{kV/m}$ and $\SI{0.4}{kV/m}$, respectively. In contrast, when the reflector is connected to photocathode potential, the field strength along the envelope glass near the photocathode layer varies only between $\sim \SI{5}{V/m}$ and $\sim \SI{25}{V/m}$.

\begin{figure}[tb]
    \centering
    \includegraphics[scale=1.0]{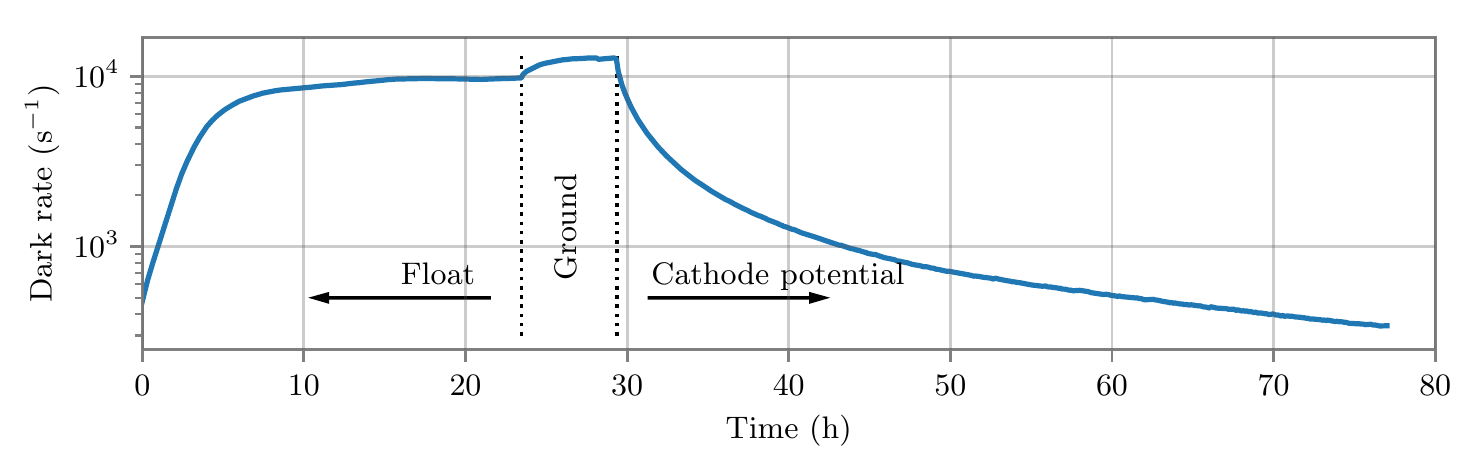}
    \caption{Dark rate as a function of time of a PMT with a reflector near its photocathode (see Fig.~\ref{fig:dr:comsol}). The reflector's potential is stepwise set at floating, ground and photocathode potential. The error bars on the rate are smaller than the line width.}
    \label{fig:dr:reflectorlong}
\end{figure}

\begin{figure}[tb]
    \centering
  \includegraphics[scale=1.0]{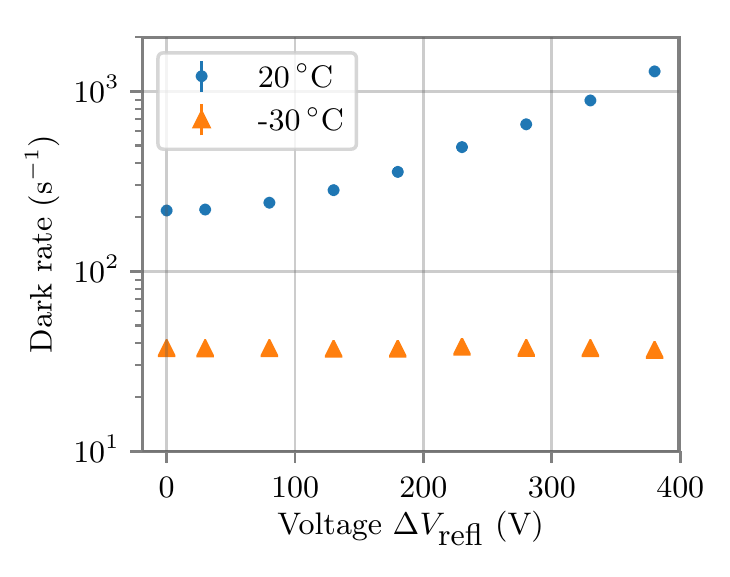}
    \caption{Dark rate of a PMT with reflector as a function of potential difference between the reflector and the photocathode at $\SI{20}{\celsius}$ (blue circles) and $\SI{-30}{\celsius}$ (orange triangles). The error bars are smaller than the data markers.}
    \label{fig:dr:reflectordep}
\end{figure}

The influence of the reflector potential on the dark rates is shown in Fig.~\ref{fig:dr:reflectorlong}. While the reflector is at floating potential, the dark rate increases slowly to a plateau of $\SI{9600}{s^{-1}}$. This process takes a very long time ($\sim\SI{15}{\hour}$) compared to the rate variation in the last section which was almost instantaneous upon changing the potential of the coating (see Fig.~\ref{fig:dr:coatingratechangetime} right). Afterwards, the reflector was connected to ground which increased the rate to a second plateau at $\SI{12800}{s^{-1}}$. Finally, the reflector was connected to the photocathode potential, which lead to a slow reduction in rate. After $\sim\SI{45}{\hour}$, the dark rate reaches a value comparable to the one of the PMT without the reflector ring.

\begin{figure}[t]
\includegraphics[width=0.495\textwidth]{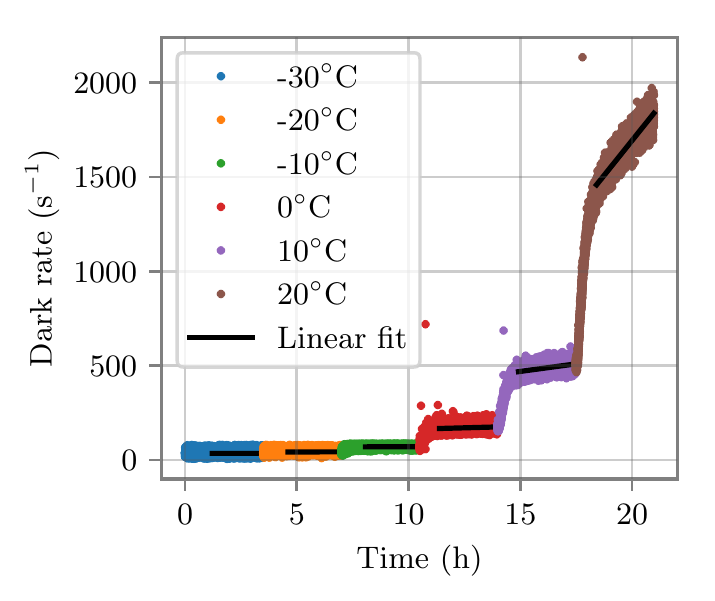}
\hfill
\includegraphics[width=0.495\textwidth]{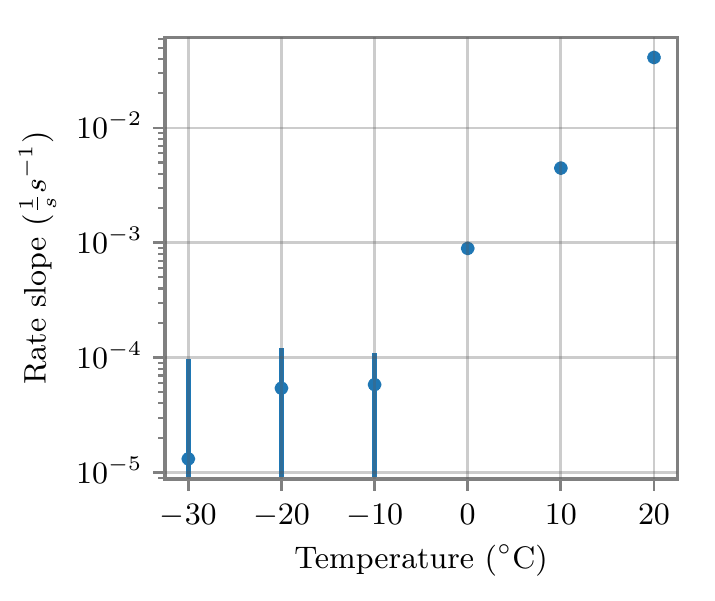}

\caption{\emph{Left:} Dark rate of a PMT as a function of temperature with a reflector at $\Delta V_{\textrm{refl}} = +\SI{400}{V}$ with respect to the photocathode. The rate increase at each temperature was fitted with a linear function shown as black lines. \emph{Right:} Slope of the linear fit as a function of temperature.}
\label{fig:dr:reflectortemp}
\end{figure}

In order to investigate this effect more systematically, the rates at different potential differences between photocathode and reflector $\Delta V_{\textrm{refl}}$ were measured for $3$ hours. Figure~\ref{fig:dr:reflectordep} shows the average rate during the last hour. This was done once at room temperature ($\SI{20}{\celsius}$) and once at $\SI{-30}{\celsius}$. At $\SI{20}{\celsius}$, the rate increases with growing $\Delta V_{\textrm{refl}}$ as expected. However, in contrast to the results from Fig.~\ref{fig:dr:coatingratechangetime}, at $\SI{-30}{\celsius}$ the rate remains constant throughout the measurement. In order to further explore this behaviour, the dark rate was measured at temperatures from $\SI{-30}{\celsius}$ to $\SI{20}{\celsius}$ at a constant potential difference of $\Delta V_{\textrm{refl}} = \SI{400}{V}$. The raw data is presented in the left plot of Fig.~\ref{fig:dr:reflectortemp}. At low temperatures, there is almost no noticeable change of the dark rate. With the stepwise increase of the temperature, the dark rate raises abruptly due to the thermal emission of electrons from the photocathode. At $\SI{0}{\celsius}$, $\SI{10}{\celsius}$ and $\SI{20}{\celsius}$ the rate clearly increases further due to the influence of the reflector after the temperature stabilises. By fitting the dark rate with a linear function (only data $>\SI{30}{\min}$ after the temperature change was used), we obtain the slope of the increase (right plot of Fig.~\ref{fig:dr:reflectortemp}). It is noticeable that the slope increases with temperature. Even at $\SI{-30}{\celsius}$ there is a positive slope, although between $\SI{-30}{\celsius}$ and $\SI{-10}{\celsius}$ it is compatible with a constant rate taking into account the uncertainties. This would explain the unchanged rate measured at $\SI{-30}{\celsius}$ in Fig.~\ref{fig:dr:reflectordep}, since it was measured only for $\SI{3}{\hour}$. With a rate increase of about $10^{-5}\,\rm{s^{-1} / s}$, one would expect a rise of only $\SI{\sim 0.1}{s^{-1}}$ after this time.

The slow increase and decrease of the rate indicates that this effect is different to the one presented in the section on the HA coating. Also, the temperature behaviour can neither be explained by the luminescence hypothesis nor by discharges between the reflector and PMT. Indeed, it has been reported that the PMT photocathode is very sensitive to electric field gradients, reducing the PMT's sensitivity over time \cite{Lavoie1967}. This has been explained as a damage of the photocathode through poisoning by the migration of sodium ions from the envelope glass to the bi-alkali layer. This could also explain the results of our measurements. The change in dark rate may be caused by ions transported to the photocathode. Ions at the glass boundary would slowly build up an opposing field causing a stabilisation of the dark rate as observed in Fig.~\ref{fig:dr:reflectorlong}. According to \cite{Mehrer2008}, the ion transport is directly proportional to the temperature. Thus, it is expected that ions will drift slower at lower temperatures resulting in a slower rise of the dark rate. But independent of the exact cause, our measurements clearly show that reflectors should be operated as close as possible to photocathode potential in order to avoid high dark rates.

\subsection{Probability of correlated pulses}
Depending on the generation process and the time offset with respect to a regular pulse, different types of correlated background pulses occur. For consistency purposes, this work uses the nomenclature and time intervals defined in \cite{km3netPaper} unless stated otherwise:
\begin{itemize}
    \item \textbf{Prepulses} are produced by photons releasing electrons from the first dynode instead of the photocathode. The resulting pulse arrives earlier than expected from the main peak of the transit time distribution. In this work we considered pulses arriving inside a time window between $\SI{60.5}{ns}$ and $\SI{5}{ns}$ prior to the regular signal. This is a more conservative range than the one used in \cite{km3netPaper} of $\SI{-60.5}{ns}$ to $\SI{-10.5}{ns}$ .
    \item \textbf{Delayed pulses} are the result of photoelectrons back-scattered elastically from the first dynode without producing secondary electrons. After re-acceleration they may start a regular multiplication process with the final pulse arriving later than regular pulses. As in \cite{km3netPaper}, we consider pulses delayed by $\SI{15.5}{ns}$ to $\SI{60.5}{ns}$.
     \item \textbf{Afterpulses type I} are produced by light emission from the dynodes and span the time window between $\SI{10}{ns}$ to $\SI{80}{ns}$ after an initial regular pulse.
    \item \textbf{Afterpulses type II} are caused by ionisation of residual gas atoms by photoelectrons. While the photoelectrons continue, producing a pulse, the positively charged ion is accelerated towards the photocathode, releasing electrons upon impact on the bialkali layer. The arrival times of pulses produced by this process range from hundreds of nanoseconds to several microseconds, depending on the average drift velocity of the ions. The time interval is chosen as in \cite{km3netPaper} between $\SI{100.5}{ns}$ and $\SI{10}{\mu s}$ after the average transit time of the PMT. 
\end{itemize}

\begin{figure}[t]
\includegraphics[scale=0.95]{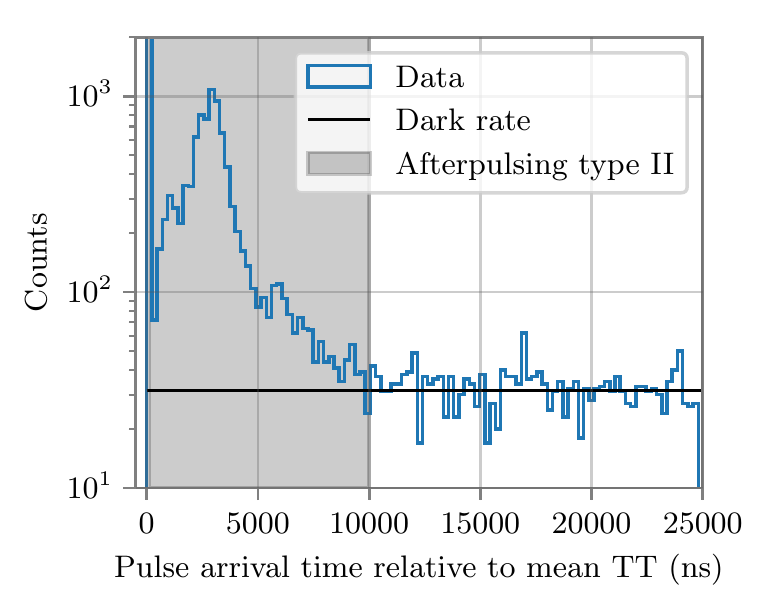}
\hfill
\includegraphics[scale=0.95]{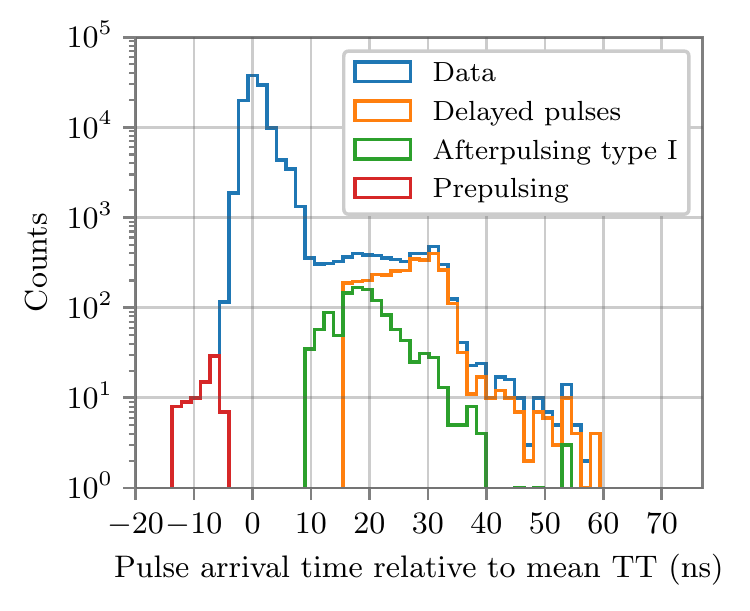}

\caption{\emph{Left:} transit-time distribution of afterpulsing type II. \emph{Right:} transit time distribution with contributions from delayed pulses, prepulses and afterpulses type I.}
\label{fig:af:tt}
\end{figure}

The measurement of these correlated pulses was performed with external triggering on an LED. The light source was set at such intensity that the average number of detected photons was $<\SI{0.1}{PE}$ at the expected transit time of the PMT. After each trigger, a waveform of $\SI{25}{\micro\second}$ was recorded and the charge, amplitude and arrival time of any pulse surpassing the $\SI{0.3}{PE}$ threshold was stored. The average TT of the PMT was calculated, fitting a Gaussian to the main TT peak, and subtracted from the arrival time of subsequent pulses. A distribution of the arrival times of such a measurement is shown in Fig.~\ref{fig:af:tt} to the left for short time intervals and to the right for the full waveforms. \mbox{Afterpulsing type II} is found up to $\sim \SI{10}{\micro\second}$ and afterwards only a constant contribution of dark rates is observed. In order to calculate the afterpulsing probability, only waveforms with a regular signal arriving in the main TT peak were considered, i.e.\ pulses between $\SI{-4}{ns}$ and $\SI{10}{ns}$. The afterpulse probability is defined as $N_{\rm{af}} / N_{\rm{p}}$, where $N_{\rm{p}}$ is the number of waveforms with a photoelectron in the main TT peak and $N_{\rm{af}}$ is the number of afterpulses in the respective time interval.
For delayed pulses only waveforms without a pulse in the main TT peak were considered. The respective probability is the ratio between the number of pulses in the time interval for delayed pulses, $N_{\textrm{d}}$, and the number of pulses with a pulse in the main TT peak $N_{\rm{p}}$. In order to calculate the prepulsing probability, the number of pulses in the prepulsing time interval was counted and divided by $N_{\rm{p}}$.

The recorded waveforms also contain pulses from uncorrelated dark noise. This artificially increases the calculated probabilities and may add a bias to the temperature dependence. To correct for this effect, the average number of dark-noise pulses $\overline{n}$ was calculated counting the number of pulses with arrival times between $\SI{15}{\mu s}$ and $\SI{25}{\mu s}$, where no afterpulsing contribution is expected (see left plot of Fig.~\ref{fig:af:tt}). Thus, the corrected number of correlated pulses $N_{\textrm{C}}$ is
\begin{equation}
    N_{\textrm{C}} = N_{\textrm{T}}-\frac{\overline{n} \cdot \Delta t}{10 \mu s},
\end{equation}
where $N_{\textrm{T}}$ is the total number of correlated pulses in the time interval $\Delta t$.

\begin{figure}[p]
    \centering
    \includegraphics[width=\textwidth]{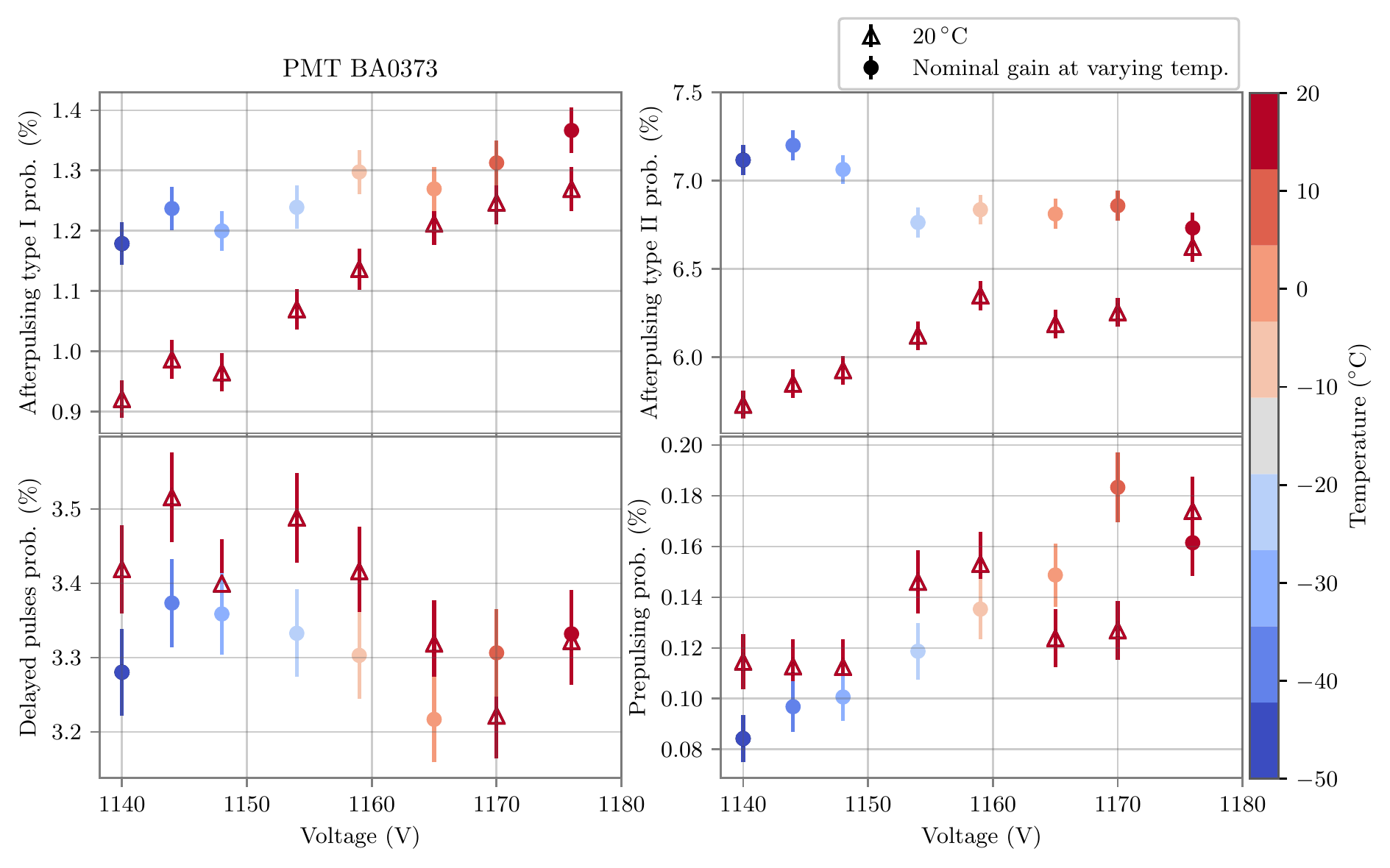}\\
    \includegraphics[width=\textwidth]{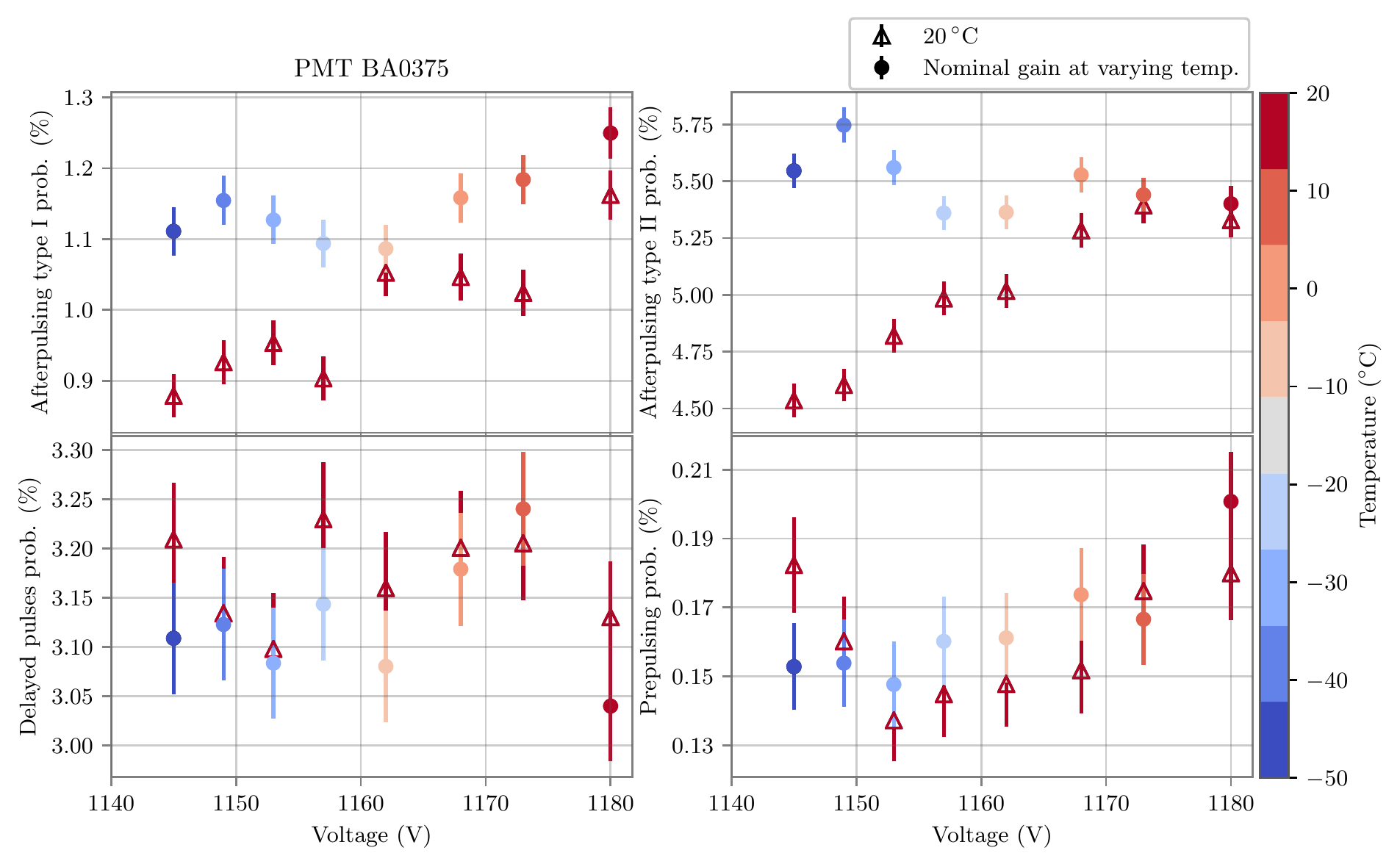}
    \caption{Probability of afterpulsing type I and II, delayed pulses and prepulses as a function of voltage. The colour of the markers indicate the temperature. Measurements symbolised by open triangles were taken at a constant temperature of $\SI{20}{\celsius}$ and varying voltages. Full circles mark measurements taken at the nominal voltage of $5 \times 10^6$ at different temperatures. The results for PMT BA0373 are shown in the top four plots and for PMT BA0375 in the bottom four plots.}
    \label{fig:dr:afterpulsing_temp}
\end{figure}

The probabilities for the different types of correlated pulses were measured and calculated for temperatures between $\SI{-50}{\celsius}$ and $\SI{20}{\celsius}$. The voltage was adjusted to keep the PMT at nominal gain. Since afterpulsing increases with voltage \cite{wrightgaintemp}, the measurement was repeated at the same voltages but at a constant temperature of $\SI{20}{\celsius}$. In this case, the PMT does not operate at constant gain. The results are presented in Fig.~\ref{fig:dr:afterpulsing_temp}. At $\SI{20}{\celsius}$ and nominal voltage the PMTs have an \mbox{afterpulsing type I} probability of $\SI{1.3}{\percent}$ and $\SI{1.2}{\percent}$ for the PMT BA0373 and BA0375, respectively. Reducing the voltage but keeping the temperature constant, the probability decreases to $\sim \SI{0.9}{\percent}$ for both PMTs (a relative reduction of approximately $\SI{28}{\percent}$ and $\SI{22}{\percent}$, respectively). A similar behaviour is found for the probability of \mbox{afterpulsing type II}. At nominal voltage the probability is $\SI{6.6}{\percent}$ (BA373) and $\SI{5.3}{\percent}$ (BA0375), respectively. At the lowest voltage tested these values are approximately $\SI{16}{\percent}$ smaller. However, measurements at different temperatures show a less steep decrease of \mbox{afterpulsing type I} than expected from just a voltage reduction, with a $\sim \SI{10}{\percent}$ relative reduction between the lowest and highest voltage. This means that effectively, there is an increase of the \mbox{afterpulsing type I} probability with decreasing temperature. This is also the case for the probability of \mbox{afterpulsing type II}, where the probability at $\SI{-50}{\celsius}$ is $\sim \SI{5}{\percent}$ larger than at $\SI{20}{\celsius}$.

The probability of delayed pulses shows neither a measurable temperature nor a voltage dependence. The average probability of delayed pulses taking all measurements into account is $\SI{3.35\pm0.02}{\percent}$ for PMT BA0375 and $\SI{3.15\pm0.02}{\percent}$ for BA0373. Finally, the prepulsing probability seems to slightly increase with voltage, but changing the temperature yields no noticeable change.

It should be noted that a reduction of the PMT voltage at a constant temperature results in a lower PMT gain. Since the threshold for a pulse is the same for all measurements, more SPE pulses are lost in measurements with lower gain, which may skew the results towards low probabilities. However, this effect is negligible, as can be seen in the probabilities for delayed pulses and prepulsing: In both cases, the results at low temperatures (where the PMT is always at nominal gain) and at room temperature (where the gain varies) are essentially the same.

\begin{figure}[t]
\centering
\includegraphics[width=0.99\textwidth]{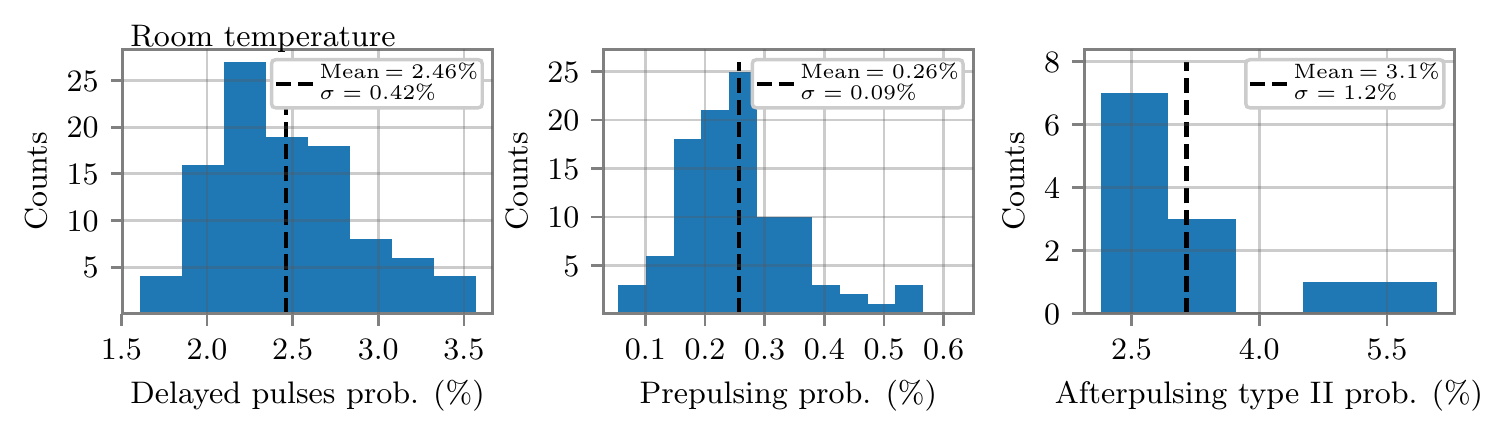}
\caption{\emph{Left and center}: Probability of delayed pulses and prepulses of $102$ PMTs at room temperature. \emph{Right}: Probability of \mbox{afterpulses of type II} at room temperature of a subset of $12$ PMTs with a threshold of $0.5\,$PE.}
\label{fig:af:erlangen_pre_del}
\end{figure}

The probabilities for delayed pulses and prepulses at room temperature of the $102$ PMTs are presented in the left and centre plot of Fig.~\ref{fig:af:erlangen_pre_del}, respectively. The average probability of delayed pulses is $\SI{2.46}{\percent}$ with a standard deviation of $\SI{0.42}{\percent}$. In the case of the prepulsing probabilities, the average is $\SI{0.26}{\percent}$ and the standard deviation $\SI{0.09}{\percent}$. This is a higher value than the average of $\SI{0.2}{\percent}$ presented in \cite{km3netPaper}. However, in this work a more conservative time interval was chosen.

Since the measurement of \mbox{afterpulses type II} takes longer, only a subset of 12 PMTs were tested at room temperature. Due to a noisy baseline, these measurements were performed with a threshold of $0.5\,$PE. The probability distribution is shown in the right plot of Fig.~\ref{fig:af:erlangen_pre_del} featuring a mean of $\SI{3.1}{\percent}$ and a standard deviation of $\SI{1.2}{\percent}$. Signal loss due to the higher trigger level is expected to be small: for PMTs BA0373 and BA0375 the afterpulsing probability changes by less than $\SI{10}{\percent}$.
%The afterpulsing probabilities of the PMTs BA0373 and BA0375 at room temperature measured with a threshold of $0.5\,$PE are only $\sim \SI{9}{\percent}$ smaller than with a threshold of $0.3\,$PE. 

\section{Conclusions}
\label{sec:conclusion}

Table~\ref{tab:concl} presents the average values of the main parameters for the R12199-01 HA MOD PMTs measured for this work at room temperature. For comparison, the parameters of the R12199-02 PMT from \cite{km3netPaper,km3netsmall} are included. As expected, the newly developed PMT model features a comparable, slightly superior performance compared to the original model. No significant degradation of the tested PMT parameters at lower temperatures was observed. As generally expected for PMTs, cooling significantly reduces the dark rate which is favourable for the use in low-background environments. As the gain was found to increase with decreasing temperatures, the nominal PMT voltage for a particular gain will need to be set accordingly. The noise-reduction measures introduced, that is the HA coating and the connection of the reflectors to photocathode potential of the respective PMTs, was found to be necessary and effective also at low temperatures.
Overall, the Hamamatsu R12199-01 HA MOD is well suited for low temperature applications such as potential future deep-ice neutrino detectors at the South Pole.

\begin{table}[H]
\centering
\caption{Main PMT parameters of the investigated PMT model R12199-01 HA MOD and its predecessor R12199-02 measured at the respective gain $G$ at room temperature. The numbers in brackets after the values indicate the number of tested PMTs.}
\label{tab:concl}
\begin{tabular}{@{}lllllll@{}}
\toprule
                    & \multicolumn{2}{l}{\begin{tabular}[c]{@{}l@{}}R12199-02\\  ($G=3\times10^{6}$)\end{tabular}} & \multicolumn{2}{l}{\begin{tabular}[c]{@{}l@{}}R12199-02\\  ($G=5\times10^{6}$)\end{tabular}} & \multicolumn{2}{l}{\begin{tabular}[c]{@{}l@{}}R12199-01 HA MOD\\  ($G=5\times10^{6}$)\end{tabular}} \\ \midrule
Nominal voltage     &                                              &                                              & 1183                                                     & [181]                            & 1148                                             & [102]                                           \\
TTS                 &                                              &                                              & 1.66                                                     & [210]                            & 1.49                                             & [102]                                           \\
prepulses          & 0.2\%                                        & [6960]                                       & 0.16\%                                                   & [20]                            & 0.26\%                                           & [102]                                           \\
Delayed pulses      & 3.2\%                                        & [6960]                                       & 3\%                                                      & [179]                            & 2.46\%                                           & [102]                                           \\
Afterpulses Type I  &                                              &                                              & \multirow{2}{*}{6.3\% $\dagger$}           & \multirow{2}{*}{[38]}            & 1.3\%                                            & [2]                                            \\
Afterpulses Type II & 7.1\%                                        & [6960]                                       &                                                          &                                  & 3.1\%                                            & [12]                                            \\ \bottomrule
\end{tabular}\\
\footnotesize{$\dagger$: Probability of afterpulses type I and II together.}
\end{table}

\acknowledgments

This work was supported by the German Bundesministerium f\"ur Bildung und Forschung (BMBF) Verbundforschung grants 05A17PM1 and 05A17WE2, and Deutsche Forschungsgemeinschaft (DFG) Research Training Group 2149.

\bibliographystyle{JHEP}
\bibliography{bib.bib}
\end{document}